\documentclass{article}

\usepackage{PRIMEarxiv}

\usepackage[utf8]{inputenc}
\usepackage[T1]{fontenc}
\usepackage{hyperref}
\usepackage{url}
\usepackage{fancyhdr}
\usepackage{graphicx}
\usepackage{amsmath}
\usepackage{amssymb}
\usepackage{amsfonts}
\usepackage{algorithm}
\usepackage{bm}
\usepackage{booktabs}
\usepackage{caption}
\usepackage{nameref}

\title{
A nonparametric statistical method for deconvolving densities in the analysis of proteomic data
}

\author{
Akin Anarat\textsuperscript{1,2,*},
Jean Krutmann\textsuperscript{2},
Holger Schwender\textsuperscript{1}\\
\textsuperscript{1}Mathematical Institute, Heinrich Heine University, Düsseldorf, Germany\\
\textsuperscript{2}IUF -- Leibniz Research Institute for Environmental Medicine, Düsseldorf, Germany\\
\texttt{*akin.anarat@hhu.de}
}

\begin{document}

\maketitle

\vspace{-1cm}

\begin{center}
\small
Published in \emph{BMC Bioinformatics}, 27(1):83, 2026.\\
\href{https://doi.org/10.1186/s12859-026-06408-0}
{doi:10.1186/s12859-026-06408-0}
\end{center}

\vspace{1cm}

\begin{abstract}
\textbf{Background:}
Genomic or proteomic data are frequently affected by noise or are convolutions of different biological signals. This is, e.g., particularly relevant in skin aging research, in which intrinsic aging, driven by genetic factors, and extrinsic aging, caused by environmental exposure, are both investigated considering, e.g., the proteome of skin fibroblasts. Since all skin areas are affected by intrinsic aging, isolating the pure extrinsic signal from the mixture of the intrinsic and extrinsic effects is crucial to investigate extrinsic aging. In such settings, deconvolution methods can be used to estimate the density of the target component. However, existing nonparametric deconvolution approaches often fail when the variance of the mixed distribution is substantially greater than that of the target distribution, a common issue in genomic and proteomic data.

\textbf{Results:}
We introduce a new nonparametric deconvolution method called N-Power Fourier Deconvolution (NPFD) that deals with this problem of differing variances by employing the $N$-th power of the Fourier transforms of the densities. Leveraging Fourier inversion and key properties of density transforms, NPFD reduces numerical instability, resulting in smooth and accurate density estimates. NPFD is able to effectively resolve smoothness- and variance-related challenges and performs comparably or better than existing methods in almost all considered scenarios, as a comprehensive simulation study shows. Moreover, applications to real proteomic data from skin fibroblasts demonstrate how NPFD can be employed to estimate the pure extrinsic aging signal.

\textbf{Conclusions:}
NPFD represents a new conceptual framework for nonparametric density deconvolution by exploiting the properties of Fourier transforms in a novel way. Its ability to address smoothness- and variance-related challenges makes it a versatile and powerful tool for deconvolving complex biological signals on the level of density functions across diverse applications.
\end{abstract}

\keywords{
Density deconvolution \and
Fourier inversion \and
Nonparametric density estimation \and
Proteomic data
}

\section{Introduction}\label{sec1}

In biomedical applications, deconvolving densities can help to tackle multiple challenges. Two situations are particularly relevant in this context. First, many biomedical factors cannot be accurately measured, leading to observations tainted by measurement errors and uncertainties. Second, it is often necessary to separate mixed signals to understand the underlying biological processes, as observed data frequently reflect a combination of distinct biological sources or overlapping effects. Examples include the measurement of hormone levels \cite{Veldhuis1992}, genetic expression data \cite{Kang2021}, and saturated fat consumption \cite{CaiHorrace2021}.

In the first situation, the goal is to estimate the density of the uncontaminated data based on measurements affected by random error. Such errors may arise from limitations in technology \cite{Charpignon2023}, self-reporting inaccuracies \cite{Althubaiti2016}, or biological variability \cite{Badrick2021}. For instance, dietary records of saturated fat intake or protein concentrations in proteomics are often subject to misreporting or technical noise \cite{Banna2015, Bogdanow2016}. In clinical trials, short-term blood pressure readings are used to infer long-term health outcomes, despite being affected by measurement error \cite{Wan2019}. As noted by Carroll et al. \cite{Carroll}, such errors can bias statistical analyses and obscure true relationships. 

Since the observed values are the sum of the true values and random errors, deconvolution techniques play a crucial role in correcting for these inaccuracies. By leveraging additional information, such as replicated measurements, or by incorporating assumptions about the error distribution, deconvolution methods enable more precise estimation of the true values \cite{CaiHorrace2021}.

Historically, much research on deconvolution methods was conducted under the assumption that the error distribution is known. However, in scenarios in which the magnitude of error is unknown, multiple measurements of the same characteristic can still be informative. If such measurements are taken with different instruments or at different times, they can be used both to estimate the true underlying value and to assess the extent of measurement error. Recently, deconvolution methods have been developed for such cases. These procedures estimate the error distribution by using repeated measurements to extract variance components \cite{Wang} or by employing estimators specifically designed to model the error structure \cite{DelaigleHallMeister}. There is also interest in estimating target densities when no additional data are available. However, this typically involves specific parametric models for error distributions \cite{Wang} or particular assumptions about the target distribution \cite{DelaigleHall}.

In the second situation, in which the signal of interest can only be observed in combination with another signal, the goal is to extract the pure signal of interest from convolved data. For example, gene expression levels in heterogeneous tissue samples may reflect mixtures of different cell types, requiring separation to isolate each cell type’s contribution. In their study, Alonso-Moreda et al. \cite{Alonso-Moreda2023} compared various methods for deconvolving cell mixtures and identifying gene signatures of blood, immune, and cancer cells, ultimately determining the most effective strategies for disentangling such complex signals.

Another example is the study of skin aging, where intrinsic and extrinsic aging are influenced by genetic and environmental factors, respectively. A major challenge in analyzing extrinsic aging lies in the fact that all human skin inherently undergoes intrinsic aging. As a result, measurements in areas exposed to external factors yield a mixture of both intrinsic and extrinsic signals. It is, therefore, necessary to extract the pure extrinsic signal from the mixture of extrinsic and intrinsic effects to investigate the impact of environmental factors on the aging process. In Section~\nameref{applreal}, we show how the nonparametric deconvolution method proposed in this article can be employed for this purpose in an application to the proteomic data collected in the GerontoSys study \cite{Waldera} dedicated to the analysis of different sources of skin aging.

Deconvolution techniques can thus improve the reliability of biomedical measurements and enhance the understanding of underlying biological processes. Many methods assume full knowledge of the convolving density, but in practice, this distribution is often only partially known. A more realistic setting involves having a sample from one of the convolving distributions alongside the mixed data. Diggle and Hall \cite{DiggleHall} and Neumann \cite{Neumann} developed nonparametric deconvolution methods for such situations. These approaches, including methods based on truncated Fourier inversion \cite{DiggleHall} and minimax convergence rates \cite{Neumann}, can be effective. However, they tend to perform poorly when the observed mixture distribution exhibits substantially higher variance than the signal to be recovered \cite{DiggleHall, Neumann}. This is a common situation in applications such as genomic or proteomic data, including the data from the GerontoSys study \cite{Waldera}. Additionally, difficulties arise when the convolving density is very smooth, which complicates the Fourier inversion process \cite{fan1991optimal}.

To address these limitations, we propose a new nonparametric deconvolution method called N-Power Fourier Deconvolution (NPFD). This approach leverages the $N$-th power of the Fourier transforms of densities to suppress high-frequency noise and improve the stability of Fourier inversion. By exploiting convolution properties such as the additivity of means and variances, NPFD disentangles mixed signals without relying on strong distributional assumptions. Its use of exponentiated Fourier-transformed values mitigates numerical inaccuracies, making it particularly robust in settings with large variance differences or very smooth convolving densities. While originally developed for the second situation discussed in this introduction, in which a sample from one of the convolving distributions is available, NPFD can also be applied in the first situation, making NPFD a flexible procedure that can effectively handle both cases within the same methodological approach.

The rest of this article is structured as follows. In Section \nameref{sec2}, the models considered in the two situations and their assumptions are described. In Section \nameref{NPFD}, NPFD is introduced and its handling of numerical inaccuracies in deconvolution problems is discussed. In Section \nameref{sims}, NPFD is evaluated in a comprehensive simulation study covering a variety of scenarios, including challenging cases with large differences in variance. Moreover, NPFD is compared to several established deconvolution methods in this simulation study. In Section \nameref{applreal}, we apply the proposed approach to blood pressure data from the Framingham Heart Study \cite{Carroll} as well as proteomic data from the GerontoSys study \cite{Waldera}. Section \nameref{conclusions} concludes this article with final remarks.

\section{Models and assumptions}\label{sec2}

The first situation describes scenarios in which only noisy observations of a variable of interest are available. Thus, the true signal, i.e., the true values of this variable, is convolved with measurement error. In this situation, the objective of deconvolution is to estimate the density $f_Y$ of the random variable $Y$ representing the true, error-free signal based on observed noisy data $z_1, \ldots, z_n$, which are assumed to be independent realizations of the random variable $Z$ with density $f_Z$. It is typically assumed in this situation that these observations follow an additive measurement error model
\begin{equation}\label{adderr}
Z_j = Y_j + X_j, \quad j = 1, \ldots, n,
\end{equation}
\noindent where $Y_1, \ldots, Y_n$ are independent and identically distributed (i.i.d.) random variables with density $f_Y$, and $X_1, \ldots, X_n$ are i.i.d. random variables with density $f_X$, which model the measurement error. Moreover, $X_j$ and $Y_j$, $j = 1, \ldots, n$, are also assumed to be independent.

The independence assumption is crucial in this case, as it implies the convolution relationship that underlies deconvolution. The additive model is appropriate in settings in which the mixed distribution can be decomposed into an underlying signal and an error component that is unrelated to the signal, such as technical measurement noise or instrument imprecision. If the measurement error depends on the latent signal so that $X$ and $Y$ are correlated, the convolution structure generally no longer holds and ordinary deconvolution methods, including NPFD, do not consistently recover the true density of $Y$.

In deconvolution settings, information about the error density $f_X$ is required to estimate the target density $f_Y$ accurately. One common way to obtain such information is through replicated measurements, i.e., repeated observations of the same latent signal $Y$ under independent realizations of the error \cite{DelaigleHallMeister}. The variation among these repeated observations reflects the influence of the error distribution and can provide a basis for estimating $f_X$. This, in turn, enables the application of deconvolution techniques to separate the effect of the error from the observed data, resulting in a more accurate estimation of $f_Y$.

If replicate measurements for the noisy observations are available, e.g., several measurements of the blood pressure of a person under the same condition, it becomes feasible to estimate the target density $f_Y$ without additional information on the variance of the error distribution. This holds as long as the error is centered, i.e., has zero mean. In this situation, there are $L$ replicated measurements
\begin{equation}\label{multsamp}
Z_{jl} = Y_j + X_{jl}, \quad l = 1, \ldots, L,
\end{equation} 
\noindent for each observation $j = 1, \dots, n$, where $Z_{j1}, \ldots, Z_{jL}$ represent the replicate measurements and the corresponding measurement errors $X_{j1}, \ldots, X_{jL}$ are assumed to be i.i.d. and independent of $Y_j$, $j = 1, \dots, n$.

When no replicated measurements are available, i.e., when only one data set of the distorted data is provided, it becomes necessary to make an assumption either about the error distribution or the target distribution. For instance, $f_Y$ can be effectively estimated under the assumption that it cannot be expressed as a convolution of two densities where at least one originates from a symmetric distribution, as this approach helps in accurately estimating $f_X$ \cite{DelaigleHall}. Another, more typical approach is to assume that $X_1, \dots, X_n$ follow a known distribution such as a centered normal distribution or a Laplace distribution \cite{Wang}.

In contrast to the first situation, in which estimating the true density $f_Y$ requires assumptions about the error distribution, the second situation, in which the observed data reflect a combination of two underlying sources, requires additional data from one of the convolving distributions to be available. This enables the estimation of the unknown distribution without imposing assumptions on the distributions involved \cite{DiggleHall}. Specifically, we assume in this situation that the observations $z_1, \ldots, z_{n_z}$ are drawn from a distribution with density $f_Z$, which represents a mixture of two underlying signals that arises as the convolution
\begin{equation}\label{convolution}
f_Z(t) = \bigl(f_Y*f_X\bigr)(t) = \int_{-\infty}^\infty f_Y(t-y)f_X(y) \ dy
\end{equation}
\noindent of two underlying densities. Here, $f_X$ represents the density of an additionally available sample $x_1, \ldots, x_{n_x}$, which originates from one of the two convolution components. This setting enables the estimation of $f_Y$ solely based on the observed samples from $f_Z$ and $f_X$, without requiring prior knowledge of their functional forms. The convolution equation \eqref{convolution} formalizes the relationship between the observed mixture distribution and its underlying components, forming the foundation for deconvolution methods.

Since the sum of two independent random variables, as given in equation \eqref{adderr}, is represented by the convolution of their distributions, the problems considered in both situations originate from equation \eqref{convolution}. The key difference between the two situations lies in the available information: in the first situation, assumptions about $f_X$ are required, while in the second situation, an additional sample from $f_X$ is available. Notably, the first situation can be viewed as a restricted version of the second, in which no external sample from $f_X$ is available and additional structural assumptions are needed instead. While NPFD described in detail in the following section has been developed in the context of the second situation, it can also be applied in the first situation and be easily adapted for corresponding estimation tasks.

\section{Methods}\label{NPFD}

In this section, we introduce a new deconvolution approach called N-Power Fourier Deconvolution (NPFD). This nonparametric method builds upon the general idea for deconvolving densities, which will be briefly introduced in Section~\nameref{general}. Unlike most established techniques (of which the ones most closely related to NPFD are described in Section~S2 of Supplementary file S1) that rely on smoothing functions to address numerical instabilities, NPFD applies a specific transformation to the observed data. This transformation enables a stabilization step that reduces the impact of fluctuations in parts of the deconvolution procedure that are particularly sensitive to numerical noise. This can result in a more flexible and potentially more accurate handling of the deconvolution problem, enhancing both reliability and applicability. In Section \nameref{NPT}, we discuss the theoretical framework of NPFD. In Section~\nameref{NPFDZ}, the complete NPFD procedure is presented. Technical details on the different components of NPFD can be found in Section~S3 of Supplementary file S1.

\subsection{General idea of deconvolving densities} \label{general}

Assuming that for given densities $f_X$ and $f_Z$ the convolution equation \eqref{convolution} holds, the unknown density $f_Y$ can be recovered via Fourier transforms, where the Fourier transform $\phi$ of a function $f$ is defined as 
$$
\phi(t) = \int_{-\infty}^{\infty} f(x) e^{i t x} dx.
$$ 
Since convolution in the density domain corresponds to multiplication in the frequency domain \cite{Rudin1987}, the Fourier transforms $\phi_X$, $\phi_Y$, and $\phi_Z$ of $f_X$, $f_Y$, and $f_Z$, respectively, satisfy 
$$
\phi_Z(t) = \phi_Y(t) \cdot \phi_X(t).
$$
This leads to the inversion formula
\begin{equation}\label{inversion}
    f_Y(y) = \frac{1}{2\pi} \int_{-\infty}^{\infty} \frac{\phi_Z(t)}{\phi_X(t)} e^{-ity} \, dt
\end{equation}
for determining $f_Y$. In practice, this expression often becomes numerically unstable due to estimation errors and the presence of small values of $\phi_X(t)$ in the denominator, particularly in the tails of the function.

Two well-known sources of this instability are the smoothness of the convolving distribution $f_X$ \cite{fan1991optimal} and a relatively large variance of $X$ compared to $Y$ \cite{DelaigleHall}, which results in a low signal-to-noise ratio. Both sources can cause $\vert \phi_X(t) \vert$ to attain relatively small values compared to $\vert \phi_Z(t) \vert$, which amplifies noise in the division in equation~\eqref{inversion}. A detailed discussion on how these two sources affect the stability of the deconvolution is provided in Section~S1 of Supplementary file S1.

\subsection{N-power transformation}\label{NPT}

To address the deconvolution problems associated with the smoothness of the convolving density $f_X$ and the high variance ratio between $X$ and $Y$, a suitable transformation of the data can be highly beneficial. In NPFD, a linear transformation is applied to the $n_x$ observations $x_1, \ldots, x_{n_x}$ from the convolving distribution and the $n_z$ observations $z_1, \ldots, z_{n_z}$ from the convolved distribution as a first step.

For a more detailed explanation, we consider the densities $f_X$ and $f_Z$. For a positive scaling parameter $a>0$ and real-valued shift parameters $b_x$ and $b_z$, we define
\begin{equation}\label{trafos} 
\tilde{X} := aX + b_x \qquad \textnormal{and} \qquad \tilde{Z} := aZ + b_z, 
\end{equation} 
\noindent where $a$, $b_x$, and $b_z$ can be specified as discussed in the following. 

Using the fact that the Fourier transform of the linear transformation $\tilde{X}$ of $X$ is given by
\[
\phi_{\tilde{X}}(t) = \mathrm E \big(e^{it\tilde X}\big) = e^{itb_x} \mathrm E \big(e^{itaX}\big) = e^{itb_x} \phi_X(at),
\]
and analogously for $\phi_{\tilde{Z}}(t)$, the quotient of the Fourier transforms takes the form
\begin{equation}\label{FTQ}
\frac{\phi_{\tilde{Z}}(t)}{\phi_{\tilde{X}}(t)}\, =\, \frac{e^{ib_zt}\phi_Z(at)}{e^{ib_xt}\phi_X(at)}\, =\, e^{i(b_z-b_x)t}\phi_{Y}(at)\, =\, \phi_{aY+b_y}(t).
\end{equation}
Hence, the analogous transformation of $Y$ is given by 
\begin{equation}\label{Ytrafo}
\tilde{Y} := aY + b_y \quad \text{with} \,\, b_y = b_z - b_x.
\end{equation}
Thus, with the uniqueness of Fourier transforms \cite{Rudin1987}, we obtain that $f_{\tilde{Z}} = f_{\tilde{Y}}*f_{\tilde{X}}$.

For an integer $N$, we set the scaling and shifting constants as
\begin{equation}\label{constants}
    a = \frac{1}{\sqrt{N}}, \qquad b_x = \left(\frac{1}{N} - \frac{1}{\sqrt{N}}\right)\textnormal{E}(X), \qquad b_z = \left(\frac{1}{N} - \frac{1}{\sqrt{N}}\right)\textnormal{E}(Z),
\end{equation}
\noindent where the choices of these constants are motivated by the general properties of expectation and variance under convolution. Specifically, if $Z$ is a random variable with density function $f_Z = f_Y * f_X$, then
\begin{equation}\label{EVar}
    \text{E}(Z) = \text{E}(Y) + \text{E}(X) \quad \text{and} \quad \text{Var}(Z) = \text{Var}(Y) + \text{Var}(X),
\end{equation}
even if $X$ and $Y$ are not independent (see Section~S1.2 of Supplementary file S1 for a detailed derivation), while in an additive measurement error model independence between $X$ and $Y$ is still required to interpret $f_Z$ as the density of the sum $X+Y$. Applying \eqref{EVar} to the shift parameters yields
$$
b_y\, =\, b_z - b_x\, =\, \left(\frac{1}{N}-\frac{1}{\sqrt{N}}\right)\textnormal{E}(Y).
$$

To interpret the $N$-th power in a probabilistically meaningful way, we consider $N$ independent random variables $Y_1,\ldots,Y_N$ that all follow the same distribution as $Y$. This allows us to express the $N$-th power of $\phi_{\tilde{Y}}$ as the Fourier transform of a sum of such i.i.d. variables. More specifically, by transforming these random variables as in \eqref{Ytrafo} to obtain $\tilde{Y}_1, \ldots, \tilde{Y}_N$, the $N$-th power of the ratio in \eqref{FTQ} is given by
\begin{equation}\label{fracN}
\left(\frac{\phi_{\tilde{Z}}(t)}{\phi_{\tilde{X}}(t)}\right)^N\, =\, \bigl(\phi_{\tilde{Y}}(t)\bigr)^N\, =\, \phi_{\sum_{k = 1}^N\tilde{Y}_k}(t)\, =\, \phi_{a\sum_{k = 1}^NY_k+\left(1-\sqrt{N}\right)\textnormal{E}(Y)}(t).
\end{equation}
\noindent In the second equation, we use the property that the convolution of two density functions translates into the multiplication of their Fourier transforms \cite{Rudin1987}. Thus, by linearly transforming the random variables $X$ and $Z$ as in \eqref{trafos} and taking the $N$-th power of the quotient of their Fourier transforms, we can calculate the density function of the sum of i.i.d. random variables $Y_1, \ldots, Y_N$ from the same distribution as $Y$. The resulting sum is scaled and shifted in such a way that the expected value and the variance match those of $Y$.

To clarify the role of the transformation in \eqref{trafos} and the subsequent power operation in \eqref{fracN}, we note the following. While the linear transformations in \eqref{trafos} are initially used to generate the Fourier transform~\eqref{FTQ} of $\tilde Y$, they are specifically chosen to make the later application of the $N$-th power in \eqref{fracN} meaningful in the sense that the resulting $N$-fold convolution is scaled and shifted in such a way that it retains the mean and variance of $Y$, while staying closely related to the distribution of $Y$ through its construction from i.i.d. copies of $Y$. Taking the $N$-th power in \eqref{fracN} constitutes the core idea of the NPFD procedure, as it helps to reduce numerical inaccuracies that arise particularly when the variance of $X$ is (much) larger than that of $Y$. In such a setting, the ratio $\phi_Z / \phi_X$ becomes unstable to determine when $\phi_Z$ and $\phi_X$ are replaced by empirical estimators. Without numerical inaccuracies, a simple linear inverse transformation of $\tilde{Y}$ in \eqref{FTQ} would lead to a reliable estimate of $f_Y$. However, this linear transformation alone does not address the numerical challenges in deconvolution, as it does not tackle the problem of a low signal-to-noise ratio induced by the relation $\sigma_Y^2 / \sigma_X^2$ of the variances of $Y$ and $X$. Specifically, since $\sigma_{\tilde{Y}}^2 / \sigma_{\tilde{X}}^2 = \sigma_Y^2 / \sigma_X^2$, the signal-to-noise ratio remains unchanged, while the additional application of the $N$-th power leads to computational stabilization.

The key motivation for applying the $N$-th power to the ratio of Fourier transforms lies in its stabilizing effect in the tails, which can have a large impact on the overall estimation of $f_Y$. Since the Fourier transforms of probability densities lie between $-1$ and $1$, small values in the tails can cause erratic fluctuations in the estimated ratio. Raising the quotient to the $N$-th power attenuates such fluctuations, mitigating numerical instability, while preserving the main structure of the target transform. However, overly large $N$ can lead to oversmoothing. Specifically, as $N$ increases, the centered and scaled sum
\[
a\sum_{k=1}^N Y_k+(1-\sqrt N)\textnormal{E}(Y)
\] 
from \eqref{fracN} converges in distribution (when $\textnormal{Var}(Y)<\infty$) to a normal limit distribution with mean $\textnormal{E}(Y)$ and variance $\textnormal{Var}(Y)$ by the Central Limit Theorem. Thus, $N$ should be chosen to balance stability and accuracy. A procedure for selecting $N$ is discussed in Section~\nameref{NPFDZ}.

\subsection{N-Power Fourier Deconvolution}\label{NPFDZ}

In this section, we present the NPFD procedure which makes use of the ideas discussed in the previous section. Since NPFD was originally developed for the deconvolution situation in which observations from both the mixed distribution and the convolving distribution are available, we first describe the NPFD procedure for this case, and then outline the modifications required for additive measurement error models. After presenting the full NPFD algorithm, we briefly explain the ideas motivating its more technical steps. For the sake of notational simplicity, we collect the available observations from the distributions of $X$ and $Z$ in the vectors $\bm{x} = \bigl[ x_1, \ldots, x_{n_x} \bigr]^\top$ and $\bm{z} = \bigl[ z_1, \ldots, z_{n_z} \bigr]^\top$. 

In the first step of NPFD, the vectors $\bm{x}$ and $\bm{z}$ are transformed according to \eqref{trafos} so that the transformed vectors are given by $\bm{\tilde{x}} = a\bm{x}+b_{x}$ and $\bm{\tilde{z}} = a\bm{z}+b_{z}$, where $a$, $b_x$, and $b_z$ are determined as described in Section~\nameref{NPT}, with $b_x$ and $b_z$ being estimated by employing the arithmetic means of $\bm{x}$ and $\bm{z}$, respectively. The value of $N$ is chosen as discussed below. 

Following these transformations, the densities $f_{\tilde{X}}$ and $f_{\tilde{Z}}$ of the transformed vectors are estimated in the second step of NPFD using a semiparametric density estimator based on a Poisson regression applied to histogram counts and a natural cubic spline \cite{efron1996}. Details of the density estimation can be found in Section~S3.1 of Supplementary file S1. 

The estimates $\widehat{f}_{\tilde{X}}$ and $\widehat{f}_{\tilde{Z}}$ are determined on the same interval $\bigl[u, v\bigr]$ using $\ell$ equidistant points $s_1, \ldots, s_\ell$, where $\ell$ is chosen large enough to ensure a sufficient resolution. Furthermore, the interval is specified to cover the entire range of all values in $\tilde{\bm{x}}$ and $\tilde{\bm{z}}$. Thus, $u = \min\bigl(\tilde{\bm{x}},\tilde{\bm{z}}\bigr)$ and $v = \max\bigl(\tilde{\bm{x}},\tilde{\bm{z}}\bigr)$ are employed. 

Afterwards, the Fourier transforms are estimated by numerical integration using the rectangle rule. Specifically, the estimates are determined at $K$ distinct values $t_1, \dots, t_K$ by
\begin{equation}\label{densXZ}
\begin{aligned}
\widehat{\phi}_{\tilde{X}}\bigl(t_k\bigr)\, &=\, \frac{v - u}{\ell} \sum_{j=1}^{\ell} \widehat{f}_{\tilde{X}}(s_j) \exp\bigl(is_jt_k\bigr),\\
\widehat{\phi}_{\tilde{Z}}\bigl(t_k\bigr)\, &=\, \frac{v - u}{\ell} \sum_{j=1}^{\ell} \widehat{f}_{\tilde{Z}}(s_j) \exp\bigl(is_jt_k\bigr).
\end{aligned}
\end{equation}
\noindent The values $t_1, \dots, t_K$ are chosen equidistantly in an interval that covers a suitable range of the resulting fraction of the two Fourier transforms so that $\bigl|\widehat{\phi}_{\tilde{X}}(t_k)\bigr|$ and $\bigl|\widehat{\phi}_{\tilde{Z}}(t_k)\bigr|$ do not become too small. A procedure for specifying this interval is presented below.

Moreover, for an accurate determination of $\widehat{\phi}_{\tilde{Y}}$ based on \eqref{densXZ} we choose an odd number for $K$ such that $t_{(K+1)/2} = 0$, since a Fourier transform is symmetric around zero. This is important, as this allows one to scale the estimation of the Fourier transform if necessary: For the Fourier transform $\phi$ of a density function $f$, it holds that 
\[
\phi(0) = \int f(x)e^{ix \cdot 0} \, dx = 1.
\]
However, numerical inaccuracies can cause the determination of $\widehat{\phi}_{\tilde{Y}}$ to slightly miss this property. Since an additional exponentiation with $N$ can amplify such deviations in the estimation of $\phi_{Y}$, this is particularly important when considering the $N$-th power of $\widehat{\phi}_{\tilde{Y}}$. Following \eqref{fracN}, the estimated Fourier transforms in \eqref{densXZ} can then be employed to determine the estimate $\widehat{\phi}_{\tilde{Y}}^N = \bigl(\widehat{\phi}_{\tilde{Z}} / \widehat{\phi}_{\tilde{X}}\bigr)^N$ of the $N$-th power of the Fourier transform of $f_{\tilde{Y}}$.

To obtain the NPFD estimate of $f_Y$, a suitable power $N$ can be selected to stabilize the estimated Fourier transform $\widehat{\phi}_Y$ by incrementally increasing $N$ until $\widehat{\phi}_{\tilde{Y}}^N$ barely exhibits fluctuations and its absolute values at the tails become very small, i.e., smaller than a threshold $\varepsilon > 0$. The choice of $\varepsilon$ controls the trade-off between resolution and stability and depends on the sample sizes of the observed data, i.e., on the lengths of $\bm{x}$ and $\bm{z}$. The role of the threshold value in the choice of $N$ and practical diagnostics for assessing the adequacy of $\varepsilon$ are discussed in Section~S3.3 of Supplementary file S1.

Starting with $N = 1$, we search for the smallest of the values $t_k$, $k = 1, \dots, K$, considered in \eqref{densXZ}, such that $\left|\bigl(\widehat{\phi}_{\tilde{Y}}(t_k)\bigr)^N\right| < \varepsilon$. If such a $t_k$ is found, we define $\gamma := t_k$. If no such $t_k$ is found, $N$ is incremented by one and the search is repeated. This process continues until a value $t_k$ satisfying the threshold condition is found to define $\gamma$. Because of the symmetry of Fourier transforms around $t_{(K+1)/2} = 0$, the lower bound of the interval is given by $-\gamma = t_{(K+1)/2 - R}$, if the upper bound has been specified by $\gamma = t_{(K+1)/2 + R}$ for an integer $R$. The estimate $\widehat{\phi}_{\tilde{Y}}^N$ is then determined on the symmetric interval $[-\gamma, \gamma]$ using all values $\bigl(\widehat{\phi}_{\tilde{Y}}(t_k)\bigr)^N$ with $|t_k| \leq \gamma$.

Having specified the power $N$, and thus, the interval $[-\gamma, \gamma]$, the NPFD estimate of the Fourier transform $\phi_Y$ can be determined by
$$
\widehat{\phi}_{Y}^\text{\,NPFD}(t_k)\, =\, \left(\widehat{\phi}_{\tilde{Y}}(t_k)\right)^{N}\, =\, \left(\frac{\widehat{\phi}_{\tilde{Z}}(t_k)}{\widehat{\phi}_{\tilde{X}}(t_k)}\right)^{N}, \qquad k = \frac{K+1}{2}-R, \ldots, \frac{K+1}{2}+R.
$$
To obtain the corresponding estimate of the target density $f_Y$, equidistant values $y_1, \ldots, y_{n_y}$ are chosen to span the interval from $y_1 = \min(\bm{z}) - \max(\bm{x})$ to $y_{n_y} = \max(\bm{z}) - \min(\bm{x})$, which approximates the support of a convolution of two distributions. The inverse Fourier transform of $\widehat{\phi}_Y^\text{\,NPFD}$, i.e., the NPFD estimate of $f_Y$, is then given by
\begin{equation}\label{fYNPFD}
\widehat{f}_{Y}^\text{\,NPFD}(y_m) = \frac{1}{\pi} \cdot \frac{\gamma}{K+2} \sum_{k = \frac{K+1}{2}-R}^{\frac{K+1}{2}+R} 
\widehat{\phi}_{Y}^\text{\,NPFD}(t_k) \, e^{-it_k y_m}, \qquad m = 1, \dots, n_y.
\end{equation}

As stated at the beginning of this section, the described procedure is specifically designed for deconvolution settings in which observations from both the mixed distribution and the convolving distribution are available. However, the densities $f_{\tilde{X}}$ and $f_{\tilde{Z}}$ cannot always be estimated reliably. While the density estimators yield a beneficial smoothing effect in suitable scenarios, their application becomes inappropriate when too few observations are available to obtain reasonable density estimates or an additive measurement error model is considered in which no direct sample from the convolving distribution is available. 

In the first of these two scenarios, information about $X$ is directly available through empirical observations. In contrast, in the additive measurement error model such information has to be inferred. As stated in Section~\nameref{sec2}, this can be achieved either by using replicate measurements or by assuming a parametric model for the distribution of $X$. A procedure for obtaining a sample $x_1,\ldots,x_{n_x}$ in the replicated setting is described in Section~S2.2 of Supplementary file~S1 and is used in the NPFD implementation. If a known (error) distribution for $X$ is assumed, the exact Fourier transform $\phi_{\tilde{X}}$ can be used and only $\phi_{\tilde{Z}}$ needs to be estimated from the transformed observations $\tilde{\bm{z}}$.

In these three situations, in which information about $X$ is either obtained from direct observations, reconstructed from replicate measurements, or specified through its known distribution, skipping the density estimation step of the NPFD procedure described above and working directly with the (estimated) Fourier transforms of $X$ and $Z$ has proven to be an effective choice. For this purpose, the empirical Fourier transform is used, if a sample of observations is available, and the exact transform is employed, when considering a parametric error model (for a definition of the empirical Fourier transform, see Section~S2.3 of Supplementary file~S1). The Fourier transforms are, then, evaluated at the $K$ frequency points $t_1,\ldots,t_K$ introduced earlier in this section. Subsequently, NPFD continues as described above with the selection of the power $N$.

The NPFD procedure for estimating the density $f_Y$ based on observed data subject to convolution is summarized in Algorithm~\ref{algtext}.

\begin{algorithm}[t]
\caption{The NPFD procedure}\label{algtext}
\textbf{Preprocessing:} If an additive measurement error model is considered and no direct sample from $X$ is available, obtain $\bm{x}$ from replicates or use the known transform $\phi_{\tilde{X}}$.\\
\textbf{Input:} Samples $\bm{x} = [x_1, \ldots, x_{n_x}]^\top$ and $\bm{z} = [z_1, \ldots, z_{n_z}]^\top$ from the convolving and mixed distribution, respectively.\\
Starting with $N = 1$, perform the following steps:
\begin{enumerate}
    \item \textbf{Transformation:} Compute the transformed vectors 
    $$
    \bm{\tilde{x}} = a\bm{x} + b_x \quad \text{and} \quad \bm{\tilde{z}} = a\bm{z} + b_z,
    $$
    where $a$, $b_x$ and $b_z$ are determined as described.
    \item \textbf{Density estimation:} Compute density estimates $\widehat{f}_{\tilde{X}}(s_j)$ and $\widehat{f}_{\tilde{Z}}(s_j)$ for equidistantly chosen values $s_j$, $j = 1, \dots, \ell$, using the density estimator of Efron and Tibshirani \cite{efron1996}, if sufficient numbers of observations are available for proper density estimations.
    \item \textbf{Fourier transformation:} If densities were estimated in Step~(2), compute $\widehat{\phi}_{\tilde{X}}(t_k)$ and $\widehat{\phi}_{\tilde{Z}}(t_k)$ via numerical integration over equidistant values $t_k$, $k = 1, \dots, K$. Otherwise, i.e., if the sample sizes are too small for reliable density estimations or an additive measurement error model is considered, use the empirical Fourier transform, if a corresponding sample is available, and the known transform $\phi_{\tilde{X}}$ when the convolving distribution is known.
    \item \textbf{Power selection:} Determine whether the chosen power $N$ is sufficiently large using the procedure described above and presented in detail in Algorithm~S1 of Supplementary file S1. If this is the case, proceed to Step~(5). Otherwise, increase $N$ by 1 and return to Step~(1).
    \item \textbf{Fourier inversion:} Using this value of $N$, specify $\gamma$ and determine
    $$
    \widehat{\phi}_{Y}^\text{\,NPFD}(t_k) = \left(\widehat{\phi}_{\tilde{Z}}(t_k)\bigl/\widehat{\phi}_{\tilde{X}}(t_k)\right)^N
    $$ 
    based on the interval $[-\gamma, \gamma]$. Determine the Fourier inversion of $\widehat{\phi}_{Y}^\text{\,NPFD}$ to obtain the NPFD estimate \eqref{fYNPFD}.
\end{enumerate}
\end{algorithm}

\subsubsection{Additional remarks on the NPFD procedure}\label{NPFD_additional_remarks}

In this section, we provide some further comments on relevant aspects of the NPFD procedure. 

As discussed in Section \nameref{NPT}, the Fourier transform $\phi$ of a density satisfies $|\phi(t)| \le 1$ for all real $t$. In practice, the estimated transform $\widehat{\phi}_{\tilde{Y}}(t_k)$ attains values smaller than 1 for those frequencies $t_k$ at which $\widehat{\phi}_{\tilde{X}}(t_k)$ has not yet decayed to small magnitudes. For any $t_k$ with $|\widehat{\phi}_{\tilde{Y}}(t_k)|<1$, the function $u \mapsto |u|^N$ monotonically shrinks the magnitude as $N$ increases, while values close to 1 (near $t=0$) are affected much less. Due to the specific choice of the initial linear transformations of $\bm{x}$ and $\bm{z}$ in \eqref{trafos}, the resulting powered transform $\bigl(\widehat{\phi}_{\tilde{Y}}(t_k)\bigr)^N$ remains closely aligned with the shape of the target transform $\phi_Y$ (see Section~\nameref{NPT}). This is the mechanism by which the $N$-th power attenuates tail fluctuations in $\widehat{\phi}_{\tilde{Y}}$.

While this tail attenuation motivates the use of powers $N>1$, it also imposes a constraint on how large $N$ can be chosen. As outlined in Section~\nameref{NPFDZ}, the power $N$ should not be increased arbitrarily, since overly large values induce oversmoothing because of the convergence of the true Fourier transform $\phi_{\tilde{Y}}^N$ towards the Fourier transform of a normal distribution as $N\to\infty$. In particular, when empirical Fourier transforms are used in NPFD, it is shown in Section~S3.5 of Supplementary file S1 that the estimator $\widehat{\phi}_{\tilde{Y}}^N$ converges to the Fourier transform of the $N(\widehat\mu_Y,\widehat\sigma_Y^2)$ distribution as $N\to\infty$, where $\widehat\mu_Y = \bar{\bm{z}} - \bar{\bm{x}}$ and $\widehat\sigma_Y^2 = \widehat\sigma_Z^2 - \widehat\sigma_X^2$. This asymptotic behavior does not generally carry over to NPFD considering estimated densities, as the preliminary smoothing induced by the density estimation step based on natural cubic splines changes the limit, while improving the performance of the method for moderate $N$. In Figures S1 and S2 in Section~S4.1 of Supplementary file S1, the behavior of NPFD is visualized in the Fourier domain across different values of the power $N$ for both variants of NPFD in an example.

In typical settings, the NPFD procedure selects only moderate values of $N$, unless the variances of $X$ and $Z$ are nearly identical or the observed data are extremely noisy. Nevertheless, even small choices $N>1$ already lead to a substantial stabilization of the tails of the corresponding Fourier transform (see Section~S4.1 of Supplementary file S1). Therefore, once a threshold crossing $\,|\bigl(\widehat{\phi}_{\tilde{Y}}(t)\bigr)^N|<\varepsilon\,$ is observed at $t=\gamma$, values of $|t|$ larger than $\gamma$ are expected to remain below $\varepsilon$, which motivates restricting the inversion to $[-\gamma,\gamma]$.

Besides the truncation threshold $\varepsilon$, the NPFD procedure involves several additional parameters that affect its numerical behavior. The frequency values $t_1, \dots, t_K$ control the resolution at which the Fourier domain is evaluated. The spatial values $s_1, \dots, s_{\ell}$ affect the accuracy of the numerical integration in the density-based variant of NPFD. An upper bound $N_{\max}$ is introduced in the automatic selection of the power $N$, preventing the choice of excessively large values, whereas the degrees of freedom of the natural cubic spline in the density estimator regulate the smoothness of $\widehat{f}_{\tilde{X}}$ and $\widehat{f}_{\tilde{Z}}$.

In the following sections, the NPFD algorithm is applied to both simulated and real data to evaluate its applicability to different deconvolution scenarios and to compare it with other established deconvolution methods, considering default values for the tuning parameters specified in the simulation study section.

\section{Simulation study}\label{sims}

To evaluate the performance of NPFD, we considered several simulation studies for each of the two situations discussed in Section \nameref{sec2}, i.e., scenarios in which an additive measurement error model is considered and scenarios in which data for both the mixed distribution and one of the convolving distributions are available. 

Moreover, we compared the performance of NPFD with several other deconvolution methods. More specifically, we considered for the first situation the procedure proposed by Delaigle et al. \cite{DelaigleHallMeister} and the method developed by Wang and Wang \cite{Wang}. While both methods were designed for deconvolution in additive error models, the approach by Delaigle et al. \cite{DelaigleHallMeister} relies on replicated measurements, whereas the method of Wang and Wang \cite{Wang} assumes a known error distribution. We further compared NPFD with the procedures of Diggle and Hall \cite{DiggleHall} as well as of Neumann \cite{Neumann} that were designed for the deconvolution setting in which data from one of the convolving distributions were available. In this case, the method of Diggle and Hall \cite{DiggleHall}, similar to NPFD, does not make any assumptions about the involved distributions, whereas the approach by Neumann \cite{Neumann} relies on specific prior information about the convolved distribution. A more detailed description of these four established deconvolution methods can be found in Section~S2 of Supplementary file S1.

Since all of these methods have different assumptions and consider different deconvolution settings, we employed different simulation scenarios for the comparisons with the respective procedures, focusing on scenarios considered by the different authors in the evaluation of their method and closely related scenarios with lower variance ratios between the target distribution and the convolving distribution.

Since the results of the evaluations and comparisons are similar across all sets of simulation scenarios, we exemplarily present in the following sections the detailed results of the evaluation of NPFD and its comparison with the deconvolution methods proposed by Delaigle et al. \cite{DelaigleHallMeister} and Diggle and Hall \cite{DiggleHall}, respectively. The results of the evaluation of NPFD in the other simulation scenarios and the comparisons with the deconvolution procedures proposed by Wang and Wang \cite{Wang} and by Neumann \cite{Neumann}, respectively, are summarized afterwards and discussed in detail in Section~S4 of Supplementary file S1.

\subsection{Simulation setup}

For every setting in this simulation study, we generated 500 simulated data sets. For the application of NPFD, the same values of the parameters of NPFD were employed for each simulated data set in every simulation setting, where for most settings, i.e. if not otherwise stated, the following predefined values or settings were used. In situations in which an additive measurement error model was considered, empirical Fourier transforms were used. Similarly, when data for both the mixed distribution and one of the convolving distributions were available with sample sizes $n_x, n_z \le 200$, empirical Fourier transforms were also applied. If the sample sizes in this second situation were larger, i.e., $n_x, n_z > 200$, the density estimates $\hat{f}_{\tilde{X}}$ and $\hat{f}_{\tilde{Z}}$ were evaluated at $\ell = 100$ equidistant points using the density estimator proposed by Efron and Tibshirani \cite{efron1996} for the respective simulated data sets, where, as frequently considered in applications of this density estimator, five degrees of freedom have been used for the natural cubic spline employed in this density estimation. Furthermore, the threshold value $\varepsilon$ introduced in Section \nameref{NPFDZ} has been set to $0.001$ and $N_{\max}$ was chosen as $100$.

As Nghiem and Potgieter \cite{Nghiem}, we used the integrated squared error 
$$
\int \bigl(\widehat{f}_Y(y)-f_Y(y) \bigr)^2 \, dy
$$
(ISE) and computed $10 \times \textnormal{ISE}$ in each setting to evaluate estimation accuracy. The results are visualized using box plots in which we excluded extreme outliers, since extreme errors, e.g., from nearly identical empirical variances of $X$ and $Z$, can distort the display. For a visual comparison of the estimated densities with the true densities, we selected the simulated data set with $10 \times \textnormal{ISE}$ closest to the median and plotted the densities estimated on this data set.

Accordingly, quantitative comparisons and uncertainty assessment are based on the distribution of $10\times\textnormal{ISE}$ across the 500 simulated data sets, whereas the displayed density curves serve as an illustrative visualization of a representative replicate. When considering such representative curves, it should be kept in mind that different deconvolution methods might handle specific realizations of the data differently, so that density estimates can look visually quite different even when their corresponding values of $10\times\textnormal{ISE}$ are similar. Small but systematic differences between methods are, therefore, assessed by shifts in the distribution of $10\times\textnormal{ISE}$ across replicates rather than through visual comparison of individual density estimates.

\subsection{Deconvolution in additive measurement error models with replicated data}\label{DelaigleHallMeisterComp}

As stated in the introduction of this article, it is more common that replicated measurements of the data from the mixed distribution are available than that the exact error density is known. In such situations, a sample of the error distribution can be generated without prior knowledge as long as the error distribution is symmetric around zero \cite{DelaigleHallMeister}. Once the data $\widehat{x}_j$, $j = 1, \dots, n$, used to estimate the density $f_X$ of the error distribution are generated by
\begin{equation*}
\widehat{x}_j = \frac{z_{j1} - z_{j2}}{\sqrt{2}}, \quad j = 1, \ldots, n,
\end{equation*}
where $z_{j1}$ and $z_{j2}$, $j = 1, \dots, n$, are independent pairs of replicated measurements from the convolved distribution (for details, see Section~S2.2 of Supplementary file S1), deconvolution can be applied to these data. Here, we will focus on scenarios in which two replicates of the mixed data are available. For more replicates, which would improve the accuracy of the estimation of the target density, an analogous approach can be followed.

\begin{table}[t]
\small
\renewcommand{\arraystretch}{1.4}
\centering
\resizebox{\textwidth}{!}{
\begin{tabular}{cccccccccc}
\toprule
\textbf{Sce.} & $\sigma_Y^2/\sigma_X^2$ & $n$ & $\widehat{f}_Y^{\text{\,NPFD}}$ & $\widehat{f}_Y^{\text{\,RMD}}$ &
\textbf{Sce.} & $\sigma_Y^2/\sigma_X^2$ & $n$ & $\widehat{f}_Y^{\text{\,NPFD}}$ & $\widehat{f}_Y^{\text{\,RMD}}$ \\
\midrule
\textbf{1} & 5 & 500 & 0.29 & 0.22 & \textbf{3} & 1 & 500 & 0.02 & 0.04 \\
 & &  & [0.24, 0.36] & [0.18, 0.26] & & & & [0.01, 0.03] & [0.03, 0.06] \\
\textbf{2} & 1 & 500 & 0.49 & 0.61 & \textbf{4} & 4 & 500 & 0.05 & 1.34 \\
 & &  & [0.41, 0.59] & [0.52, 0.71] & & & & [0.03, 0.06] & [1.28, 1.41] \\
\bottomrule
\end{tabular}}
\caption{The median and (in squared brackets) the first and third quartiles of $10 \times \textnormal{ISE}$ of the NPFD density estimator $\widehat{f}_Y^{\text{\,NPFD}}$ and the RMD density estimator $\widehat{f}_Y^{\text{\,RMD}}$ from 500 simulations.}
\label{tab2}
\end{table}

In order to evaluate the performance of NPFD in scenarios with replicated data, we considered in Scenario 1 a $\chi_3^2\bigl/\sqrt{6}$ distribution for the target density $f_Y$ and a $N(0, 0.2)$ distribution for the convolving density $f_X$. This scenario is based on the setting used by Delaigle et al. \cite{DelaigleHallMeister}, who considered a $\chi_3^2$ distribution as the target density. We additionally applied a scaling factor to obtain a standardized version with unit variance.

Building on the construction as in this Scenario 1, we expanded our analysis by considering additional simulation scenarios to further evaluate the performance of NPFD under varying conditions. In Scenario 2, we considered the same target density $f_Y$ as in Scenario 1, but a normal distribution with a larger variance, specifically a $N(0,1)$ distribution, for $f_X$. In Scenario 3, $f_Y$ was modeled by a $\text{Gamma}\left(12, \sqrt{3}\right)$ distribution, while the convolving density $f_X$ was drawn from a $N(0, 4)$ distribution. For Scenario 4, $f_Y$ was considered as the convolution $\chi_{1.5}^2 * N(0, 1)$ of a $\chi_{1.5}^2$ distribution with a standard normal distribution, while $f_X$ was again modeled as a $N(0, 1)$ distribution. 

The parameters of these distributions were chosen such that the ratios $\sigma_Y^2/\sigma_X^2$ of variances have round values. The sample size was set to $n = 500$. The observations $z_{j1}$ and $z_{j2}$, $j = 1, \dots, n$, for $f_Z$ were independently drawn from the same respective distribution.

Based on the practical diagnostic described in Section~S3.3 of Supplementary file S1, we employed in Scenarios 1 and 2 a value $\varepsilon = 0.1$ for the threshold used in the automatic power selection step of NPFD. The remaining parameters of the considered situations were left at their default settings described at the beginning of this section. Following the approach by Delaigle et al. \cite{DelaigleHallMeister}, which we refer to as repeated measurement deconvolution (RMD) in the following, we adjusted negative values of the density estimation resulting from NPFD by setting them to zero.

\begin{figure}[H]
\centering
\includegraphics[width=0.9\textwidth]{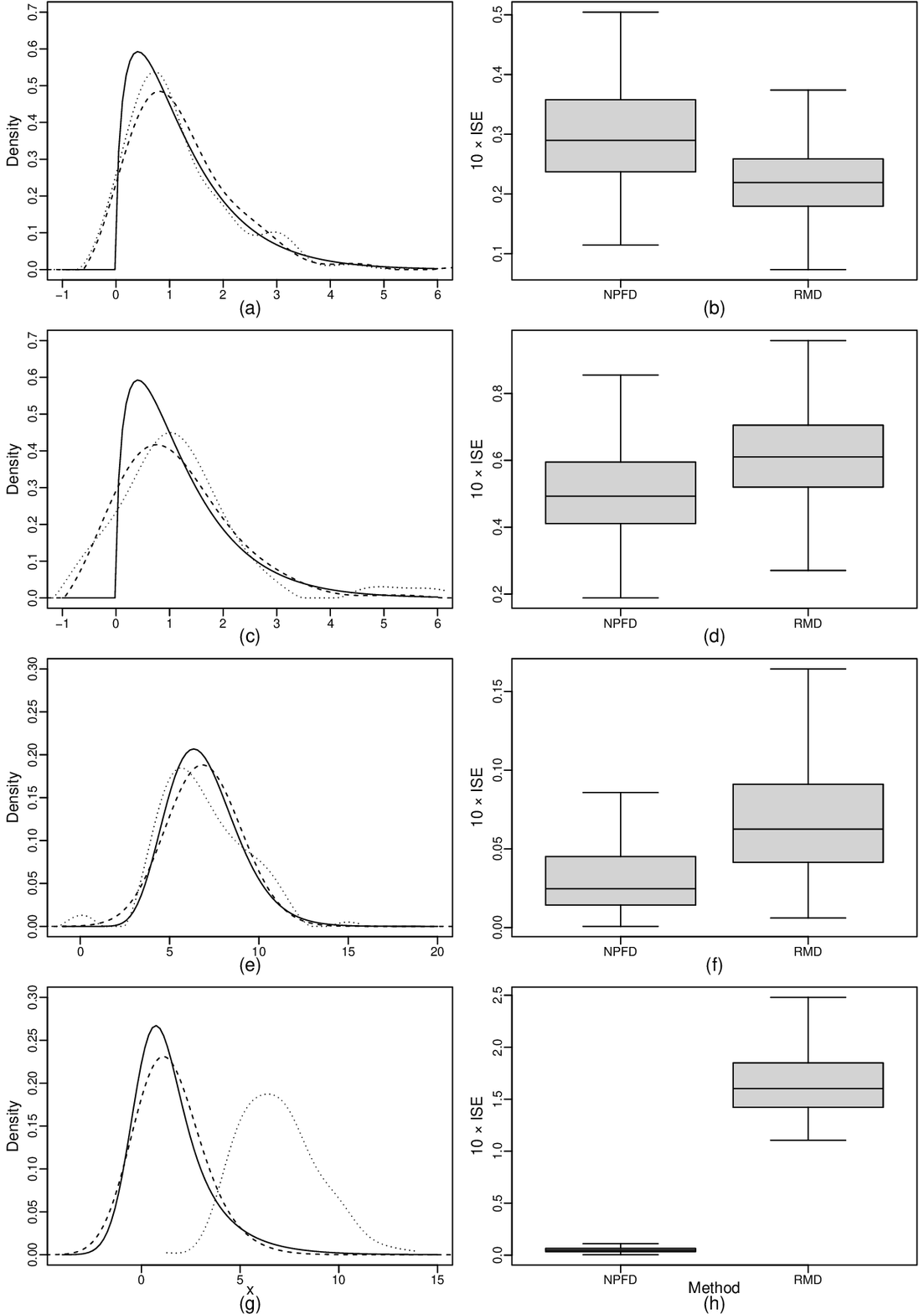}
\captionsetup{format=plain}
\caption{(a), (c), (e), (g): Comparison of a representative density estimate $\widehat{f}_Y^{\text{\,NPFD}}$ (dashed) to a representative density estimate $\widehat{f}_Y^{\text{\,RMD}}$ (dotted) with the true density $f_Y$ (solid) for Scenario 1 (a), Scenario 2 (c), Scenario 3 (e), and Scenario 4 (g). \, (b), (d), (f), (h): Box plots of the values of $10 \times \textnormal{ISE}$ of the density estimators (without outliers) in the corresponding 500 simulated data sets from (a), (c), (e), and (g), respectively.}\label{fig1}
\end{figure}

In Table \ref{tab2}, summarizing statistics for the value of $10\times \text{ISE}$ in the applications of NPFD and RMD are presented. While RMD performed better than NPFD in Scenario 1, in which a high target to error variance ratio was considered, NPFD excelled in Scenario 2 in which this ratio was substantially reduced. NPFD also outperformed the RMD method in Scenario 3, in which a more uniform target density was of interest. In Scenario 4, a substantial difference in the amount of error between NPFD and RMD is observed. A possible explanation for this is given in the following discussion of the visualized density estimates visualized in Fig. \ref{fig1}.

Fig. \ref{fig1} highlights that the density estimates produced by NPFD perform well not only in terms of accuracy, but also in terms of smoothness. Fig. \ref{fig1} (a), (b), that show the results for Scenario 1, demonstrate that despite the superior performance of RMD in this scenario, NPFD still produced a reasonable estimate of the target density by utilizing a power $N = 2$. This is noteworthy, given that the target density exhibits a sharp decline near zero on the $x$-axis, while remaining strictly positive, which complicates the deconvolution process for NPFD, as the target density changes its shape markedly under convolution with itself. However, the results of Scenario 2  displayed in Fig. \ref{fig1} (c), (d) illustrate the advantage of NPFD, when the variance ratio $\sigma_Y^2/\sigma_X^2$ is decreased. With an increased power of $N = 3$, NPFD provides in this setting superior performance over the RMD approach in terms of smoothness and accuracy. Furthermore, Fig.~\ref{fig1} (e) illustrates that the application of both NPFD and RMD resulted in good estimates for a smoother target density in Scenario 3, without a sharp decline near zero, where $f_Y^\text{\,NPFD}$ with a determined power of $N = 7$ showed a better estimation of the mode, resulting in smaller errors (see Fig. \ref{fig1} (e), (f)). In Fig. \ref{fig1} (g), (h), the limitations of RMD when dealing with a density that is a convolution with a symmetric density are visualized. The symmetric part of the target density appears to be incorrectly identified as part of the convolving error density, leading to a shift in the estimation, which is an issue that is also discussed in \cite{DelaigleHall}. In contrast, NPFD with a power of $N = 8$ was able to accurately estimate the target density in this situation.

\subsection{Deconvolution with an available sample from the convolving distribution}\label{DiggleHallComp}

In the comparisons of NPFD with the deconvolution procedure of Diggle and Hall \cite{DiggleHall}, we adopted their original simulation examples and then reduced the variance ratios $\sigma_Y^2/\sigma_X^2$. Diggle and Hall \cite{DiggleHall} examined different scenarios that involved varying convolving densities $f_X$ and varying sample sizes $n = 100$ or $n = 500$ from both the convolving density $f_X$ and the mixed density $f_Z$. They focused on estimating the density $f_Y$ of a $\textnormal{Gamma}(4, 1)$ distribution using their estimator, which we will denote FDD (Fourier deconvolution with damping) in the following. 

Building upon these settings, we considered eight simulation scenarios. In Scenarios 1–4, $f_Y$ was modeled by a Gamma(4,1) distribution, where in Scenario 1 an Exp(0.5) distribution and in Scenario 2 an Exp(0.25) distribution were considered for $f_X$, yielding higher variance of $X$ in Scenario 2 than in Scenario 1. Similarly, in Scenarios 3 and 4, a Gamma(4,2) and a Gamma(4,1) distribution, respectively, were used to simulate $f_X$. In Scenarios 5 and 6, $f_X$ followed a Weibull(4, 12.44) distribution and $f_Y$ was modeled by a $\chi^2$ distribution with 3 or 8 degrees of freedom, respectively. Finally, in Scenarios 7 and 8, a $\text{Gumbel}(-12, \sqrt{6}/\pi)$ distribution was used for $f_Y$, while $f_X$ was modeled by a $N(9,1)$ or a $N(9,2)$ distribution, both being super smooth functions, which makes the deconvolution problem more challenging (for a discussion of smooth and super smooth functions, see Section~S1.1 of Supplementary file S1). The observations for $f_X$ were drawn from the respective distribution, and the observations for $f_Z$ from the convolution $f_Z = f_Y * f_X$.

For each of the eight simulation scenarios, we generated data sets with, on the one hand, $n = n_x = n_z = 100$, and on the other hand, $n = 500$ observations. For the estimation of $f_X$ in Scenarios~1 and~2, three degrees of freedom were used for the natural cubic spline to adequately capture the shape of the exponential distribution.

In Table \ref{tab1}, the median as well as the first and third quartile of the 500 values of $10 \times \textnormal{ISE}$ are presented for the applications of NPFD and FDD to the data of each of the eight scenarios. This table shows that NPFD performed better than FDD across all considered scenarios. As the variance of the convolving distribution increases, FDD tends to produce much larger errors, whereas NPFD maintains relatively small ones. As sample sizes decrease, errors in the application of NPFD increase across all cases, while errors of FDD tend to decrease in Scenarios 6, 7, and 8. This trend can be attributed to the increasing uniformity of the empirical Fourier transforms of the mixed and convolving densities with larger sample sizes. As the results in Scenarios 7 and 8 show, NPFD can appropriately deal with super smooth convolving densities, whereas FDD struggles with such densities, which was also mentioned by Diggle and Hall \cite{DiggleHall}.

\begin{table}[H]
\small
\renewcommand{\arraystretch}{1.4}
\centering
\scalebox{0.88}{
\begin{tabular}{cccccccccc}
\toprule
\textbf{Sce.} & $\sigma_Y^2/\sigma_X^2$ & $n$ & $\widehat{f}_Y^{\text{\,NPFD}}$ & $\widehat{f}_Y^{\text{\,FDD}}$ &
\textbf{Sce.} & $\sigma_Y^2/\sigma_X^2$ & $n$ & $\widehat{f}_Y^{\text{\,NPFD}}$ & $\widehat{f}_Y^{\text{\,FDD}}$ \\
\midrule
\textbf{1} & 1 & 500 & 0.03 & 0.07 & \textbf{5} & 1 & 500 & 0.01 & 0.05 \\
 &  &  & [0.02, 0.04] & [0.05, 0.09] &  &  &  & [0.00, 0.01] & [0.03, 0.10] \\
 &  & 100 & 0.11 & 0.18 &  &  & 100 & 0.02 & 0.14 \\
 &  &  & [0.06, 0.18] & [0.11, 0.33] &  &  &  & [0.01, 0.05] & [0.07, 0.34] \\
\addlinespace
\textbf{2} & 0.25 & 500 & 0.06 & 0.14 & \textbf{6} & 0.5 & 500 & 0.02 & 1.28 \\
 &  &  & [0.04, 0.08] & [0.10, 0.24] &  &  &  & [0.01, 0.04] & [0.47, 3.70] \\
 &  & 100 & 0.31 & 0.39 &  &  & 100 & 0.05 & 1.10 \\
 &  &  & [0.17, 0.83] & [0.24, 0.81] &  &  &  & [0.02, 0.14] & [0.33, 3.29] \\
\addlinespace
\textbf{3} & 4 & 500 & 0.03 & 0.07 & \textbf{7} & 1 & 500 & 0.12 & 0.61 \\
 &  &  & [0.02, 0.04] & [0.05, 0.12] &  &  &  & [0.08, 0.19] & [0.35, 2.19] \\
 &  & 100 & 0.06 & 0.14 &  &  & 100 & 0.23 & 0.75 \\
 &  &  & [0.03, 0.09] & [0.09, 0.36] &  &  &  & [0.14, 0.37] & [0.45, 2.34] \\
\addlinespace
\textbf{4} & 1 & 500 & 0.05 & 0.47 & \textbf{8} & 0.5 & 500 & 0.21 & 1.39 \\
 &  &  & [0.03, 0.08] & [0.20, 1.53] &  &  &  & [0.13, 0.34] & [0.67, 4.14] \\
 &  & 100 & 0.12 & 0.58 &  &  & 100 & 0.35 & 1.25 \\
 &  &  & [0.07, 0.21] & [0.26, 1.99] &  &  &  & [0.21, 0.58] & [0.73, 4.05] \\
\bottomrule
\end{tabular}}
\captionsetup{format=plain}
\caption{The median and (in squared brackets) the first and third quartiles of $10 \times \textnormal{ISE}$ of the NPFD density estimator $\widehat{f}_Y^{\text{\,NPFD}}$ and the FDD density estimator $\widehat{f}_Y^{\text{\,FDD}}$ from 500 simulations with sample size $n = n_x = n_z$.}
\label{tab1}
\end{table}

\begin{figure}[H]
\centering
\includegraphics[width=0.9\textwidth]{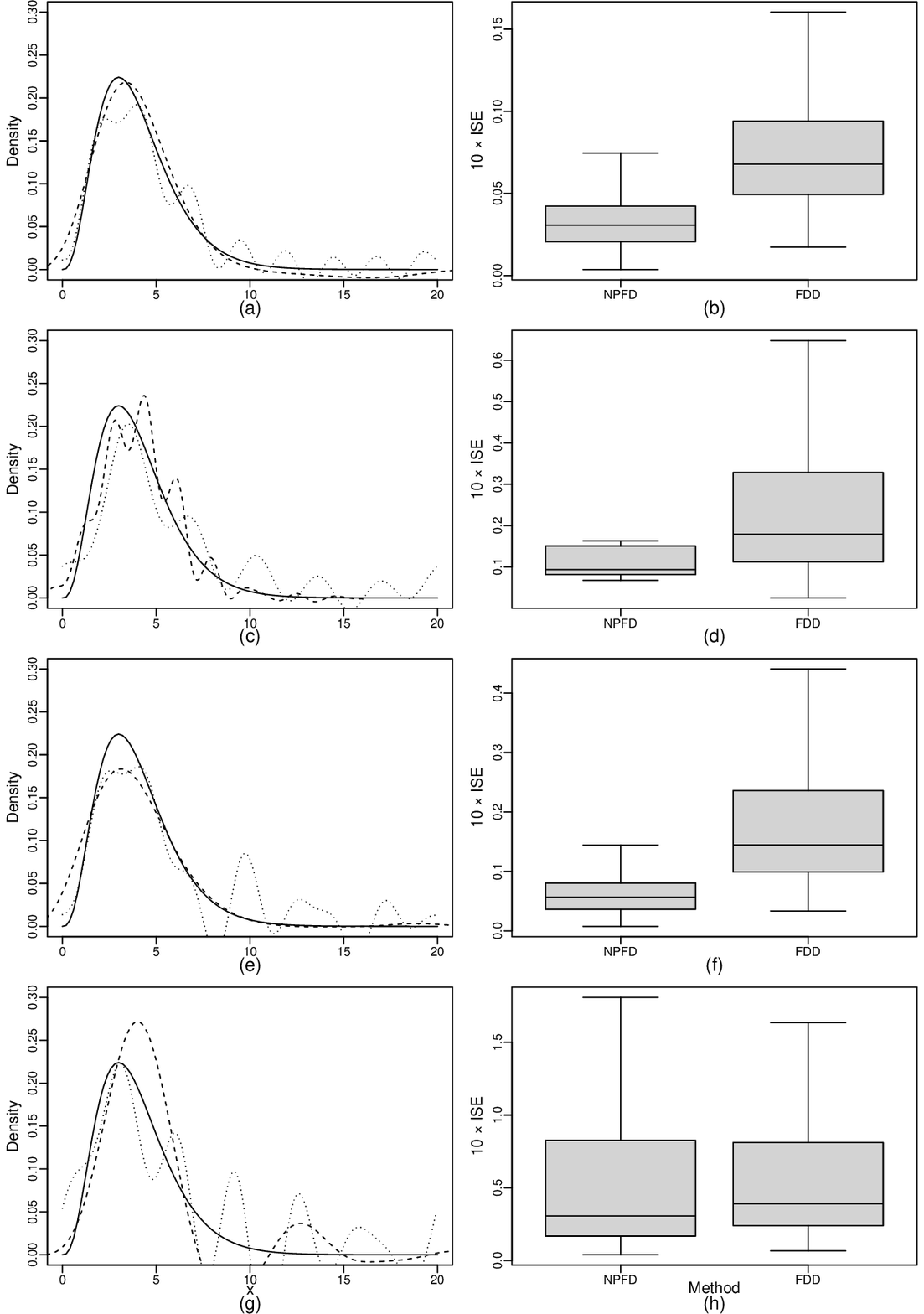}
\captionsetup{format=plain}
\caption{(a), (c), (e), (g): Comparison of a representative density estimate $\widehat{f}_Y^{\text{\,NPFD}}$ (dashed) to a representative density estimate $\widehat{f}_Y^{\text{\,FDD}}$ (dotted) with the true density $f_Y$ (solid) for Scenario 1 with sample sizes of 500 (a) and 100 (c), as well as for Scenario 2 with sample sizes of 500 (e) and 100 (g). \, (b), (d), (f), (h): Box plots of the values of $10 \times \textnormal{ISE}$ of the density estimators (without outliers) in the corresponding 500 simulated data sets from (a), (c), (e), and (g), respectively.}\label{fig2} 
\end{figure}

In Fig. \ref{fig2}, box plots of the values of $10 \times \textnormal{ISE}$ and the estimated densities of the estimators with the respective median value of $10 \times \textnormal{ISE}$ are shown for the respective two settings in Scenarios 1 and 2 to illustrate the impact of sample size and variance differences. The results of the remaining scenarios are displayed in Supplementary Figures~S3-S5. In the first setting, utilizing a power of $N = 2$, NPFD produced a much smoother density estimate compared to FDD, which provided a density estimate with a noticeably oscillatory behavior (see Fig. \ref{fig2} (a)). In addition, $f_Y$ was estimated by NPFD with noticeably lower error values than by FDD (see Fig. \ref{fig2} (b)). In the second setting shown in this figure, in which the sample sizes are decreased from 500 to 100 to examine the impact of reducing sample sizes, NPFD still generated a reliable density estimation (see Fig. \ref{fig2} (c), (d)), where a power of $N = 3$ was used in this deconvolution. NPFD, again, clearly outperformed FDD regarding smoothness and accuracy of the estimated target density. In the third displayed setting, which corresponds to Scenario 2 with a large sample size, the variance of the convolving distribution was increased to investigate the effect of a decreasing ratio $\sigma_Y^2/\sigma_X^2$. Using $N = 2$ in this setting, NPFD again achieved a promising estimation (see Fig. \ref{fig2} (e), (f)). Moreover, this setting revealed even greater differences in error magnitude and smoothness between NPFD and FDD compared to the first two settings. A similar pattern to the first three settings can be observed in Scenarios 3-8 (see Section~S4.2 of Supplementary file S1). When considering the smaller sample sizes in Scenario 2, the performance of both methods was roughly comparable (see Fig. \ref{fig2} (g), (h)). While the results of NPFD, with a power of $N = 4$, exhibited a slightly lower median error and a smaller lower quartile than FDD, they also showed a higher upper quartile regarding the deviation of the estimates from the true density. Overall, both methods estimated the true density reasonably well, where the weaker performance compared to the results from the other settings can be attributed to the combination of low sample sizes and a small variance ratio $\sigma_Y^2/\sigma_X^2$.

\subsection{Summary of comparisons with other deconvolution methods}\label{OthersComp}

In addition to the discussed comparisons with the method of Delaigle et al. \cite{DelaigleHallMeister} and the approach by Diggle and Hall \cite{DiggleHall}, we evaluated the performance of NPFD in further simulation scenarios reflecting additional deconvolution settings and compared NPFD in these settings with, on the one hand, the method of Wang and Wang \cite{Wang}, and on the other hand, the procedure
of Neumann \cite{Neumann}. Detailed descriptions of these scenarios and detailed discussions of the comparisons of the estimators can be found in Sections~S4.3 and S4.4 of Supplementary file S1, with the results displayed in Supplementary Figures~S6-S10. 

In all considered scenarios, NPFD was either comparable to or outperformed the respective competitor, despite in part requiring less prior information. Specifically, in additive measurement error models with known error distributions, NPFD yielded better results than the approach by Wang and Wang \cite{Wang} and performed particularly well when the variance of the error distribution was high. Moreover, NPFD also showed strong performance in settings including a bimodal target density and scenarios with disjoint support. When comparing NPFD to the procedure proposed by Neumann \cite{Neumann}, NPFD produced lower estimation errors, especially in scenarios with larger sample sizes or increasing variance of the convolving density.

\section{Application to real data}\label{applreal}

In the following, we apply NPFD to real data. A first application involving replicated measurements in an additive error model from the Framingham Heart Study \cite{Carroll} is presented in the following section. Afterwards, the application of NPFD to the proteomic data from the GerontoSys study \cite{Waldera} is discussed focusing on the deconvolution of these proteome data.

\subsection{Application to blood pressure data from the Framingham Heart Study}\label{framingham}

In the Framingham Heart Study, systolic blood pressure was measured twice during each of two examinations in 1615 participants, yielding four different variables representing the measurements at the two examinations, where each of the variables includes substantial measurement errors. The goal was to estimate the underlying density of systolic blood pressure corrected for measurement error.

Following Carroll et al. \cite{Carroll}, we calculated for each of the 1615 patients the average of the two systolic blood pressure measurements taken during each of the two examinations. As Wang and Wang \cite{Wang}, who also applied their deconvolution method to this data set, we used these averages to estimate the density of the error-free data at the follow-up examination, whereas the data from the first examination were used solely to estimate the error density. We accordingly applied NPFD to these data to evaluate whether NPFD is in real-world applications able to adequately deconvolve densities in an additive measurement error setting. For comparison, we also applied the deconvolution methods of Wang and Wang \cite{Wang} as well as Delaigle et al. \cite{DelaigleHallMeister}.

As in the application of the method of Delaigle et al. \cite{DelaigleHallMeister}, we only utilized in NPFD the provided data sets assuming a centered error distribution for the deconvolution. In contrast, Wang and Wang \cite{Wang} additionally assumed in their application a normal distribution for the error and used the given observations to estimate the variance of the error distribution. 
 
\begin{figure}[t]
\centering
\includegraphics[width=0.75\textwidth]{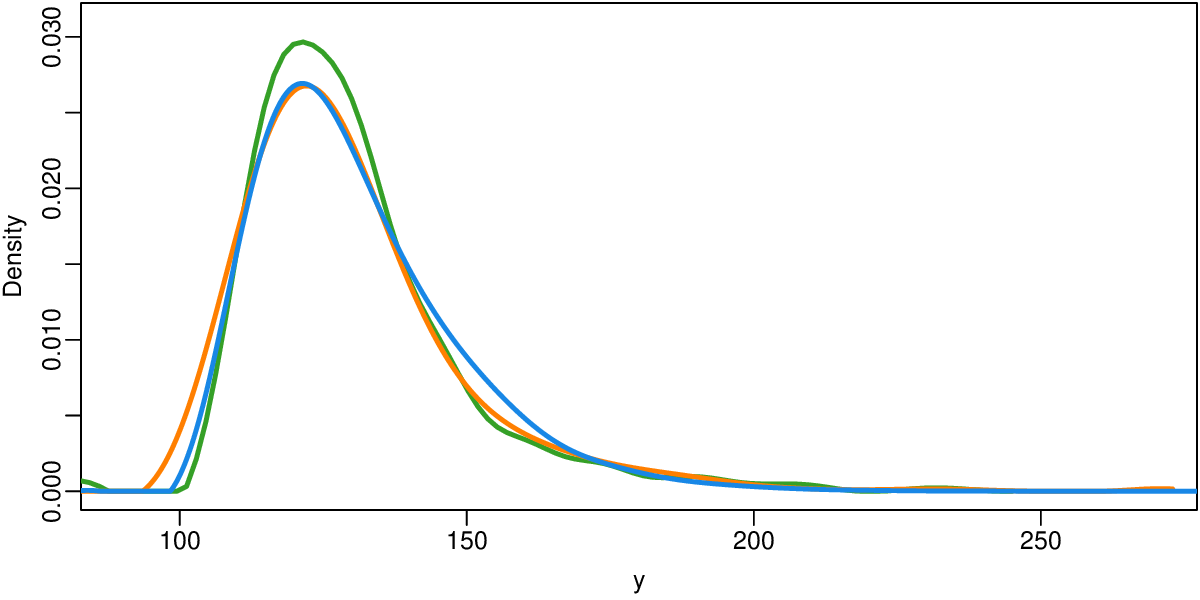}
\captionsetup{format=plain}
\caption{Deconvolution of the target density for the data of the Framingham Heart Study. The density estimator $\widehat{f}_Y^{\text{\,NPFD}}$ (blue) and the estimators proposed by Wang and Wang (orange) and Delaigle et al. (green) are shown.}\label{fig3}
\end{figure}

In Fig.~\ref{fig3}, the resulting density estimates are presented. All three methods yield similar results, where NPFD closely matches the smooth curves estimated using the two comparison procedures. Only at the mode, located identically by all three methods, the estimate of the procedure proposed by Delaigle et al. \cite{DelaigleHallMeister} slightly deviates from the other two, while NPFD and the method of Wang and Wang \cite{Wang} produced nearly identical results. This application thus shows that NPFD is able to provide an estimate of the deconvolved density in a real data situation that is similar to estimates obtained by established methods. An advantage of NPFD over other established deconvolution methods is exemplified in the following application.

\subsection{Application to proteome data from the GerontoSys study}\label{gerontosys}

The aging of skin is affected by both genetic and environmental factors, leading to intrinsic and extrinsic aging processes. To investigate the impact of these two types of factors on skin aging, on the one hand, skin areas that are not (or only very weakly) exposed to environmental factors such as UV radiation, and on the other hand, skin areas that are exposed to UV radiation and other environmental factors are considered in studies such as the GerontoSys study \cite{Waldera}. For this investigation on a molecular level, e.g., the proteome of fibroblasts in these skin areas has been measured in the GerontoSys study for 15 female donors from three age groups (18–25, 35–49, and 60–67 years; five per group). Further details on sample preparation and data acquisition in this study are provided in \cite{Waldera}.

A major challenge in the investigation of extrinsic skin aging is that all skin undergoes intrinsic aging so that areas exposed to environmental factors reflect a combination of both intrinsic and extrinsic aging. To isolate the extrinsic component, proteomes from extrinsically aged fibroblasts are compared to those from intrinsically aged regions of the same individuals. In this case, the determined pure extrinsic signal typically does not directly reflect the absolute effect size of environmental factors on the fibroblasts, but the changes in protein expression levels caused by environmental factors. These changes can then be used to, e.g., compare their values between age groups to study how extrinsic aging manifests over time.

To extract the densities of the pure extrinsic signal in the GerontoSys study, we applied NPFD to $\log_2$-transformed proteomic data from both skin areas exposed and not exposed to environmental factors. In the deconvolution framework, the sample from the non-exposed skin of each individual represents the data stemming from the convolving density $f_X$, while the sample from the exposed skin corresponds to the convolved density $f_Z$, satisfying $f_Z = f_Y * f_X$, where $f_Y$ denotes the density of the pure extrinsic signal that needs to be estimated. In this analysis, we focused on the protein data of the ten women from the younger and the older age group. In accordance with previous analyses of the GerontoSys data, proteins not present in at least three samples within any of the four combinations of age group and skin area were excluded, yielding 2379 proteins.

As stated in equation~\eqref{EVar}, deconvolution requires that the variance of the convolved distribution is larger than the variances of its components. We, therefore, examined for each of the ten donors whether the empirical variance of the protein values from the mixed intrinsic and extrinsic signal exceeded the empirical variance of the intrinsic signal. Since this requirement was not fulfilled for one person from the older age group, we excluded this person from the further analysis.

In the analysis of the proteome data for these nine women, we denote for donor $i = 1,\ldots,n_g$ in age group $g$ (young or old) the vector consisting of the log$_2$-transformed protein levels $x_1^{ig},\ldots,x_{n_x^{ig}}^{ig}$ from the non-exposed skin area by $\bm{x}_{ig} = (x_1^{ig},\ldots,x_{n_x^{ig}}^{ig})^\top$ and the vector consisting of the log$_2$-transformed protein levels $z_1^{ig},\ldots,z_{n_z^{ig}}^{ig}$ from the exposed skin area by $\bm{z}_{ig} = (z_1^{ig},\ldots,z_{n_z^{ig}}^{ig})^\top$.

\begin{figure}[t]
\centering
\includegraphics[width=1\textwidth]{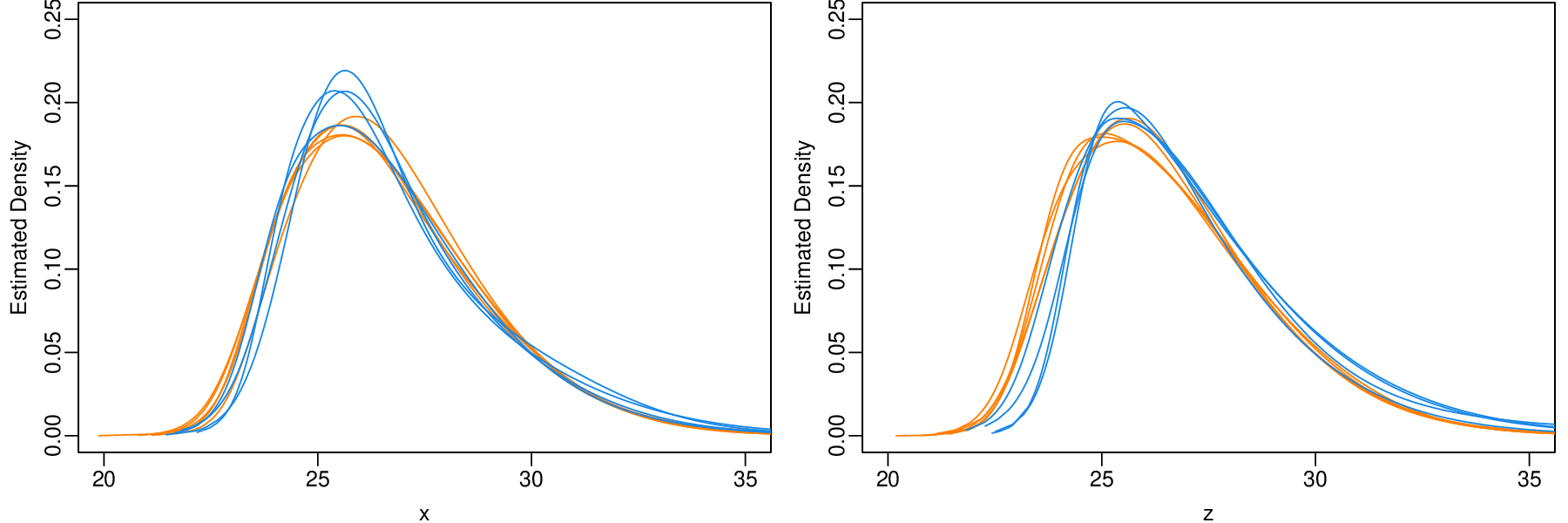}
\captionsetup{format=plain}
\caption{Estimated densities of the intrinsic signal (left panel) and the mixture of intrinsic and extrinsic signal (right panel) of the proteins from the proteome of the skin fibroblasts for the five women of the younger age group (marked by orange lines) and four women of the older age group (blue lines) from the GerontoSys study.}\label{fig4}
\end{figure}

\begin{figure}[t]
\centering
\includegraphics[width=0.75\textwidth]{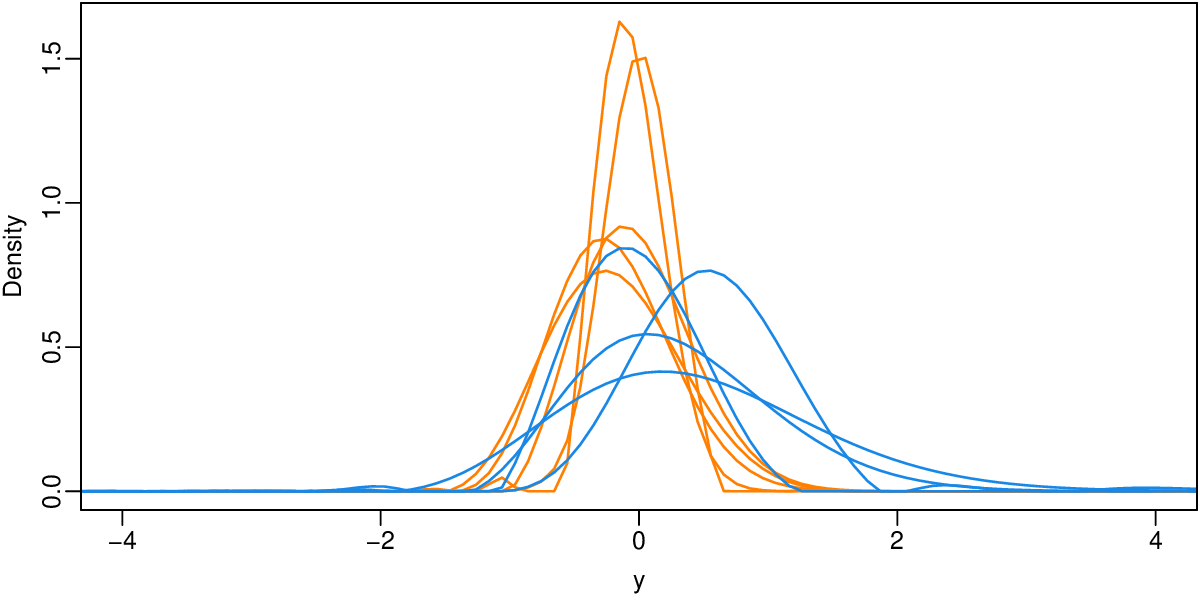}
\captionsetup{format=plain}
\caption{Deconvolved densities of the extrinsic signal of the proteins from the proteome of the skin fibroblasts for the five women of the younger age group (marked by orange lines) and four women of the older age group (blue lines) from the GerontoSys study.}\label{fig5}
\end{figure} 

To explore the signal distributions, we estimated the densities of the intrinsic signals $x_1^{ig},\ldots,x_{n_x^{ig}}^{ig}$ and the mixed signals $z_1^{ig},\ldots,z_{n_z^{ig}}^{ig}$ separately for each of the nine donors. As shown in Fig.~\ref{fig4}, the intrinsic densities are closely aligned across individuals, with slightly less variation in the older age group. In contrast, the mixed signal densities for older individuals are modestly shifted to the right, suggesting generally higher protein expression values.

To uncover the densities of the pure extrinsic signal from the mixed intrinsic and extrinsic signal of the proteins, we then applied NPFD proband-wise to the respective proteomic data. Specifically, NPFD was applied separately to each individual pair $(\bm{x}_{ig}, \bm{z}_{ig})$, treating the non-exposed measurements as observations from the convolving distribution and the exposed measurements as observations from the convolved distribution. The resulting estimates of the deconvolved densities are shown in Fig. \ref{fig5}.

NPFD yielded smooth estimates in nearly all cases. Only two of the deconvolved densities exhibit minor irregularities in small parts of their tails. Fig.~\ref{fig5} shows that for three of the four older women the deconvolved extrinsic densities are shifted to the right compared to those of the younger group, indicating stronger environmentally induced signals. These densities also exhibit greater variability, suggesting more heterogeneous environmental effects in older skin.

This visual impression is supported by the results of an application of a statistical test proposed by Delicado \cite{Delicado2007} for comparing groups of densities. This test rejects the null hypothesis that the distributions underlying the density functions of extrinsic skin aging for the younger and older female subjects are equal at a significance level of $0.05$ with a $p$-value of $0.0159$.

We also applied the method of Diggle and Hall \cite{DiggleHall} to the same data. The resulting estimates of the deconvolved densities exhibit strong oscillations and fail to represent valid probability densities (see Supplementary Figure~S11). This instability likely stems from the similar variances of the intrinsic and mixed signals (see Fig.~\ref{fig4}) and highlights the advantage of the N-power approach in NPFD in handling such variance-sensitive settings.

\section{Conclusion}\label{conclusions}

In this article, we introduced NPFD (N-Power Fourier Deconvolution), a nonparametric deconvolution method in which the $N$-th power of the estimated Fourier transform is employed to deconvolve a density from a convolution of this density with another density. Applying this N-Power transformation addresses two key challenges in deconvolution, namely a very smooth convolving distribution and large differences in the variance between the mixture and the target distribution. 

In an extensive simulation study, NPFD almost always yielded more accurate and smoother estimates than established deconvolution methods. Specifically, NPFD proved effective in additive measurement error models, performing better than methods proposed by Delaigle et al. \cite{DelaigleHallMeister} and Wang and Wang \cite{Wang}, especially when the mixture distribution had substantially higher variance than the target distribution. Furthermore, NPFD performed superiorly in the comparisons with the approaches by Diggle and Hall \cite{DiggleHall} and Neumann \cite{Neumann}, which both were developed for situations in which observations of the convolving and the convolved distribution are available.

In an application to a real data set, NPFD produced a reliable density estimate for the uncontaminated systolic blood pressure data from the Framingham Heart Study. This estimate of the deconvolved density closely resembled the estimates of the methods proposed by Delaigle et al. \cite{DelaigleHallMeister} and Wang and Wang \cite{Wang}, where these methods were, in contrast to NPFD, specifically designed to address issues such as the one represented in this real-world data scenario. 

Additionally, NPFD was applied to proteomic data from the GerontoSys study to deconvolve the density of the extrinsic signal from a mixture of the intrinsic and extrinsic signal measured in skin fibroblasts to analyze the genetic and the environmental impact on skin aging. In contrast to the application of the method proposed by Diggle and Hall \cite{DiggleHall}, which resulted in oscillatory density estimates, NPFD produced smooth densities, indicating that this procedure can effectively handle large differences in variance between the mixed and the target distribution.

In this application, we aggregated all proteins (present in the majority of the samples) to obtain one deconvolved density estimate of the extrinsic signal for each individual and to compare these distributions between the younger and the older age group. This provides a first summary of extrinsic aging effects at the distribution level and illustrates how NPFD can be used to disentangle intrinsic and extrinsic components in complex proteomic data.

However, specific biological processes or gene ontology groups may exhibit systematically different extrinsic signal distributions between age groups. A natural next step in the analysis of the GerontoSys data is, therefore, to stratify proteins by biological or gene ontology annotations and apply NPFD separately within each functional group. The resulting group-specific deconvolved densities can then be compared descriptively and, on a quantitative basis, using functional data tests \cite{Delicado2007} to identify protein groups with significantly different extrinsic signal distributions between age groups.

From a methodological perspective, it might be interesting to investigate in future research whether the performance of NPFD can be further improved by combining this procedure with other nonparametric deconvolution methods, particularly those that utilize smoothing kernels for density estimation, which have their own advantages and have proven effectiveness in various deconvolution scenarios \cite{Wang, DelaigleHallMeister}. Combining them with NPFD could enhance the overall reliability and flexibility of the deconvolution process. By leveraging the strengths of both NPFD and kernel-based approaches, it may be possible to achieve an even more accurate and efficient deconvolution.

NPFD is implemented in the \textsf{R} package \texttt{NPFD} \cite{Anarat2024}, freely available on CRAN.

\section*{Supplementary information}

Supplementary material supporting this article is available with the published version in \emph{BMC Bioinformatics}:
\href{https://doi.org/10.1186/s12859-026-06408-0}
{doi:10.1186/s12859-026-06408-0}.

\section*{Acknowledgements}

We thank Peter Kern for fruitful discussions.

\section*{Author contributions}

A.A. developed the proposed method. A.A. and H.S. drafted the main text and designed the simulation study. A.A., J.K., and H.S. conceived the analyses of the application to the proteomic data. All authors reviewed and approved the final manuscript.

\section*{Funding}

This work has been supported by the Research Training Group ``Biostatistical Methods for High-Dimensional Data in Toxicology'' (RTG 2624, Project P4) funded by the Deutsche Forschungsgemeinschaft (DFG, German Research Foundation -- Project Number 427806116).

\section*{Data availability}

The blood pressure dataset supporting the conclusions of this article is included in the \textsf{R} package \texttt{decon}, available at \url{https://www.rdocumentation.org/packages/decon/versions/1.3-4}.

The proteome dataset supporting the conclusions of this article is available from the corresponding author upon reasonable request.

\section*{Competing interests}

The authors declare no competing interests.

\end{document}